\documentclass[aps,pra,twocolumn, notitlepage,10pt,superscriptaddress,longbibliography]{revtex4-1}
\usepackage{graphicx}
\usepackage{times,bbm,amsmath,amssymb}
\usepackage{hyperref}
\usepackage{color}
\usepackage{graphicx}
\usepackage{dcolumn}
\usepackage{bm}
\usepackage{soul}

\usepackage{nicefrac, xfrac}
\usepackage{amssymb}
\usepackage{pifont}
\usepackage{tikz}

\begin{document}

\title{Deep reinforcement learning for quantum multiparameter estimation}

\author{Valeria Cimini}
\affiliation{Dipartimento di Fisica, Sapienza Universit\`{a} di Roma, Piazzale Aldo Moro 5, I-00185 Roma, Italy}

\author{Mauro Valeri}
\affiliation{Dipartimento di Fisica, Sapienza Universit\`{a} di Roma, Piazzale Aldo Moro 5, I-00185 Roma, Italy}

\author{Emanuele Polino}
\affiliation{Dipartimento di Fisica, Sapienza Universit\`{a} di Roma, Piazzale Aldo Moro 5, I-00185 Roma, Italy}

\author{Simone Piacentini}
\affiliation{Istituto di Fotonica e Nanotecnologie, Consiglio Nazionale delle Ricerche (IFN-CNR), Piazza Leonardo da Vinci, 32, I-20133 Milano, Italy}

\author{Francesco Ceccarelli}
\affiliation{Istituto di Fotonica e Nanotecnologie, Consiglio Nazionale delle Ricerche (IFN-CNR), Piazza Leonardo da Vinci, 32, I-20133 Milano, Italy}

\author{Giacomo Corrielli}
\affiliation{Istituto di Fotonica e Nanotecnologie, Consiglio Nazionale delle Ricerche (IFN-CNR), Piazza Leonardo da Vinci, 32, I-20133 Milano, Italy}

\author{Nicol\`o Spagnolo}
\affiliation{Dipartimento di Fisica, Sapienza Universit\`{a} di Roma, Piazzale Aldo Moro 5, I-00185 Roma, Italy}

\author{Roberto Osellame}
\affiliation{Istituto di Fotonica e Nanotecnologie, Consiglio Nazionale delle Ricerche (IFN-CNR), Piazza Leonardo da Vinci, 32, I-20133 Milano, Italy}

\author{Fabio Sciarrino}
\email{fabio.sciarrino@uniroma1.it}
\affiliation{Dipartimento di Fisica, Sapienza Universit\`{a} di Roma, Piazzale Aldo Moro 5, I-00185 Roma, Italy}

\begin{abstract}

Estimation of physical quantities is at the core of most scientific research and the use of quantum devices promises to enhance its performances. In real scenarios, it is fundamental to consider that the resources are limited and Bayesian adaptive estimation represents a powerful approach to efficiently allocate, during the estimation process, all the available resources.
However, this framework relies on the precise knowledge of the system model, retrieved with a fine calibration that often results computationally and experimentally demanding. 
Here, we introduce a model-free and deep learning-based approach to efficiently implement realistic Bayesian quantum metrology tasks accomplishing all the relevant challenges, without relying on any \textit{a-priori} knowledge on the system. To overcome this need, a neural network is trained directly on experimental data to learn the multiparameter Bayesian update. Then, the system is set at its optimal working point through feedbacks provided by a reinforcement learning algorithm trained to reconstruct and enhance experiment heuristics of the investigated quantum sensor. Notably, we prove experimentally the achievement of higher estimation performances than standard methods, demonstrating the strength of the combination of these two black-box algorithms on an integrated photonic circuit. This work represents an important step towards fully artificial intelligence-based quantum metrology.

\end{abstract}

\maketitle

\section*{Introduction}

Multiple-parameter estimation is an essential task for both fundamental science and applications. For this reason developing strategies and devices able to perform measurements with the smallest uncertainty has become a research branch of particular interest for various fields. It is known that the achievement of the ultimate precision bounds is possible only exploiting quantum resources \cite{giovannetti2011advances}. Indeed, nowadays, quantum sensors \cite{degen2017quantum,pirandola2018advances} represent one of the most promising applications of quantum enhanced technologies and they are already employed for different applications, from imaging \cite{brida2010experimental} and biological sensing \cite{taylor2013biological} to gravitational wave detection \cite{schnabel2010quantum,abadie2011gravitational}. The highest measurement precision achievable depends on the available resources, therefore, the main focus of most quantum metrology investigations relies on the optimization of such probe states and, successively, of the performed measurements to attain such limit \cite{polino2020photonic,barbieri2022optical}. However, in a real scenario, the number of quantum resources is always limited and not all the desired probe states can be prepared. It follows that, in a limited-resource regime, operating the device at its optimal working point and employing optimized control strategies \cite{yuan2015optimal,liu2017control} for achieving the highest estimation precision becomes crucial. The identification of such optimal feedbacks is far from being trivial in particular for quantum systems of increasing complexity and dimensions and for multiparameter estimation problems. Usually, the employed optimization algorithms are extremely time-consuming since they have to be computed after each measurement outcome and, more importantly, they rely on the knowledge of device physical model. One of the most employed methods, which assures the convergence to the ultimate precision bound, is to update the knowledge on the parameter posterior distribution through Bayes rule after the use of each resource \cite{box2011bayesian,helstrom1976quantum,li2019bayes,Valeri2020,multipara2022}. Therefore, adaptive methods generally require a precise characterization of the operation of the employed system in order to update properly the knowledge on the investigated parameters at each step of the protocol. Such requirement is still the bottleneck for the application of optimal adaptive protocols in most quantum sensing applications.
A practical model-free alternative to Bayesian update combined with a computationally feasible optimization algorithm for the identification of optimal feedbacks is thus desirable.

In this work, we simultaneously overcome these fundamental challenges by developing a deep reinforcement learning protocol which combines a  reinforcement learning (RL) agent with a deep neural network (NN), in an actual noisy multiparameter estimation experiment, where the control feedbacks are efficiently chosen by an intelligent agent that does not rely on any  explicit hardware model. All the NN trainings are performed on experimental data, therefore no additional information beside the one extracted directly from the accessible measurements is required. 
With this approach we first demonstrate the convergence to the ultimate precision bound in the single-parameter estimation scenario and then we experimentally prove to outperform standard calibration strategies, exploiting quantum resources in the limited data regime, for the simultaneous estimation of three optical phases in a state-of-the-art integrated photonic quantum sensor. 
The achievement of such good estimation performances is granted by the choice of the control feedbacks performed by a learning agent whose reward depends on the updated knowledge of the parameters after each step of the estimation protocol. Crucially, in our work the Bayesian update is obtained by a previously trained deep neural network, therefore, it does not require at any step the model of the system. With this approach, we are able to experimentally prove the validity of a  model-free optimization for parameter estimation problems, opening the way to fully artificial intelligence-based quantum metrology.  

To prove the validity of our approach we start investigating the performances of a single-parameter estimation on a test-bed system, extending the protocol developed in \cite{nolan2021machine} to the adaptive framework, training a NN for Bayesian update. We then generalize such algorithm for multiparameter estimation problems using a sequential Monte Carlo technique for the computation of the Bayesian probabilities and we combine it with a RL agent necessary to achieve good estimation precisions in more complex systems. Finally, we prove experimentally the effectiveness of the combination of a deep NN for Bayesian update with a RL agent who chooses the optimal controls on an actual multiparameter photonic quantum sensor.



\section{Artificial Intelligence quantum metrology}

Machine Learning (ML) represents a powerful alternative to the need of developing a model describing the system behaviour, for this reason its use in the most varied research fields is spreading \cite{carleo2019machine,vernuccio2022artificial}. Such techniques result particularly effective when applied to the study of quantum systems that usually live in a high-dimensional space and their characterization turns out to be a computationally hard task to solve, requiring the analysis of a huge amount of data \cite{gebhart2022learning,dawid2022modern}. Different supervised and unsupervised learning algorithms have been applied to solve efficiently quantum many-body problems\cite{carleo2017solving}, reconstruct the density matrix of high-dimensional quantum systems \cite{torlai2018neural} and even for the design of new quantum experiments \cite{melnikov2018active,dunjko2018machine,krenn2016automated,krenn2020computer,krenn2021conceptual} and the discovery of physical concepts \cite{iten2020discovering,roscher2020explainable}. Their application to the metrology and sensing fields fosters the idea of self-calibrated quantum sensors not relying on an explicit knowledge of the model describing the device operation \cite{cimini2019calibration,cimini2021calibration,nolan2021frequentist} and retrieving Hamiltonian parameters directly from experimental data \cite{gentile2021learning}. As an example, Nolan et al. in \cite{nolan2021machine} reformulated the parameter estimation problem as a classification task to overcome the calibration requirements needed from Bayesian estimation. 

The term ML refers to a huge class of algorithms sharing one common feature that is the ability of extrapolating some kind of knowledge directly from the data. These algorithms are then sub-divided into three macro-areas: supervised learning, unsupervised learning and reinforcement learning \cite{alpaydin2020introduction,sutton2018reinforcement}. While the first two kinds of algorithms have the purpose of inferring the structure relating labelled or unlabeled data, the latter refers to algorithms developed to control the dynamics of a system. This is done through a model-free feedback-based method where an intelligent agent learns to perform tasks in a defined environment depending on the reward it receives. The purpose of the agent is to find the optimal series of actions, in response to the changing state of the environment, which maximizes its reward. After the agent has been sufficiently trained it has been demonstrated that it can beat humans in several tasks, for example in playing games such as Go \cite{silver2016mastering}. When RL algorithms are combined with the output of NNs we can speak of deep reinforcement learning \cite{franccois2018introduction}.

Very recently, the use of RL algorithms is proving to be a powerful resource, indeed they have been deployed in different numerical works for quantum systems control \cite{bukov2018reinforcement,fosel2018reinforcement,niu2019universal,baum2021experimental,an2019deep}, for finding the optimal feedbacks in parameter estimation tasks enhancing the sensor dynamics \cite{xu2019generalizable,schuff2020improving,xiao2022parameter}, and for Hamiltonian learning \cite{gebhart2022learning,borah2021measurement,sivak2022model}. Interestingly, Fiderer et al. in \cite{fiderer2021neural} developed a RL method to create efficient experiment-design heuristics for Bayesian quantum estimation, gaining a great advantage over the extremely slow previously-developed algorithms \cite{hentschel2009adaptive,hentschel2011efficient}. 
However, these works still rely either on the full knowledge of the system's quantum state or on the explicit likelihood function describing the system output probabilities. Also for this reason, until now, their application has been demonstrated mostly through theoretical simulations.

\begin{figure*}[ht!]
\centering
\includegraphics[width=\textwidth]{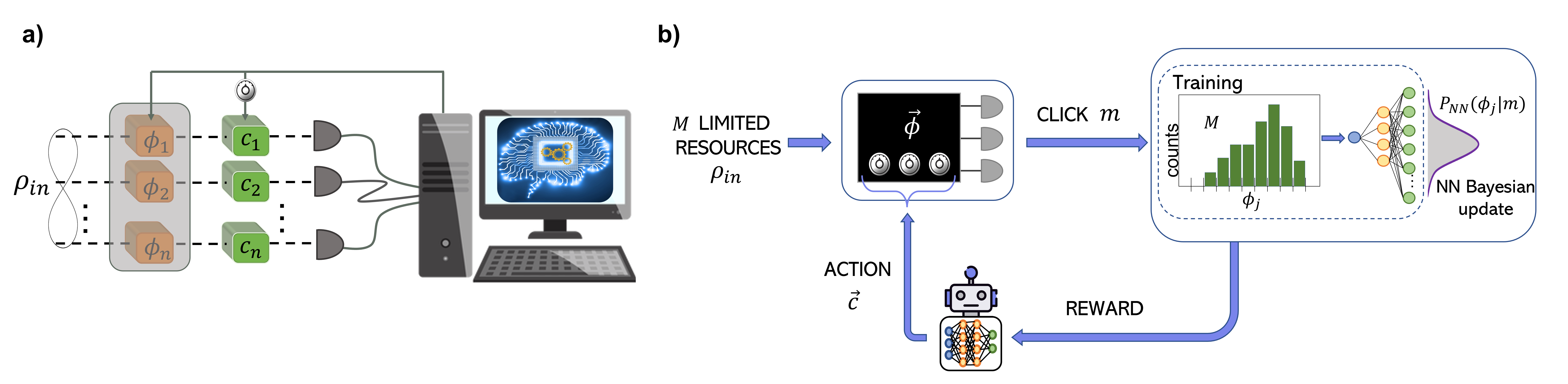}
\caption{\textbf{a)} Generic multiparameter estimation problem fully managed by artificial intelligence processes. Quantum probes evolve through the investigated system and consequently  their state changes depending on $\bm{\phi}.$ Both the single-measurement update and the setting of control parameters $\bm{c}$ is done via machine learning algorithms to optimize the information extracted per probe. \textbf{b)} Sketch of the implemented protocol. A limited number of quantum probe states are fed into the sensor treated as a black-box. A grid of measurement results is collected to train a NN which learns the posterior probability distribution associated to the single-measurement Bayesian update. Such distribution is used to define the reward of a RL agent who sets the control phases on the black-box device.}
\label{fig:protocollo}
\end{figure*}

A generalization and extension of such ML approaches is thus necessary for handling realistic metrological processes over their whole spectrum.
Here, we develop and test experimentally a protocol fully based on articificial intelligence, which governs a real noisy sensor, from the learning of how to update the Bayesian belief over the system dynamics, to the optimal choice of actions to be performed in order to speed up the estimation performances.
In order to do so, we extended and combined the two aforementioned ML algorithms \cite{nolan2021machine} and \cite{fiderer2021neural} to demonstrate a black-box adaptive multiparameter quantum estimation protocol in a real photonic device, where the unknown parameters are the relative phase shifts between the arm of an interferometer (see Fig.\ref{fig:protocollo}).

\subsection*{Neural Network Bayesian adaptive multiparameter estimation}

\subsubsection*{Bayesian learning}

The purpose of estimation protocols is to retrieve the values of a vector of parameters $\bm{\phi}$ through the measurement with a previously prepared probe. When the probe state interacts with the investigated system, its state changes depending on the parameters vector $\bm{\phi}$; therefore, the measurement of the state of the probe after such evolution allows to give an estimate of the parameters. To correctly assess their values it is necessary to reconstruct the detection probabilities of all the $d$ possible measurement outcomes. Bayesian protocols use the measurement results to update the \emph{a prior} knowledge $p(\bm{\phi})$ on the parameters under investigation retrieving the posterior probability distribution through the Bayes's rule:

\begin{equation}
    P(\bm{\phi}|\bm{d}) = \frac{P(\bm{d}|\bm{\phi})p(\bm{\phi})}{p(\bm{d})}.
\label{Bayes}    
\end{equation}
Here, $P(\bm{d}|\bm{\phi})$ is the likelihood function describing the probability of obtaining a certain measurement outcome given the set of parameters values. Knowing the explicit model of the system under study it is then possible to compute the mean of the posterior probability $ P(\bm{\phi}|\bm{d})$, reconstructed after sending $N$ probes, from which it is possible to retrieve the estimate $\bm{\hat\phi}$ of the investigated parameters as follows:

\begin{equation}
    \bm{\hat\phi} = \int \bm{\phi}  P(\bm{\phi}|\bm{d}) d\bm{\phi}.
\end{equation}

\subsubsection*{Bayesian neural network}

To overcome the need of reconstructing the explicit model of the detection probabilities, we train a feed-forward NN for the reconstruction of such posterior probability distributions. 
The network requires the discretization of the continuous parameters space in order to treat the problem as a classification task, identifying each possible value of the vector $\bm{\phi}$ as one among $N_{\bm{\phi}}$ specific labels $\bm{\phi}_1,\ldots,\bm{\phi}_{N_{\bm{\phi}}}$. The training is performed associating the single-measurement results, corresponding to the $d$ possible outcomes, to the respective label associated to the setup parameters. For each class, a fixed number of measurement repetitions $r$ has to be shown to the NN during the training allowing it to learn the correct conditional probability distribution $P_{\text{NN}}(\bm{\phi}_j|\bm{d})$.

Following the arguments of Nolan et al. in \cite{nolan2021machine}, the output of the trained NN corresponds to the Bayesian posterior distribution for each measurement outcome up to a normalization factor which depends on the parameters grid spacing. In our case such spacing results to be $\delta\phi=\frac{L}{N_{\phi}-1}$, where $L$ is the width of the interval of the parameters values. From the retrieved posterior distribution, it is possible to compute the \emph{a prior} distribution of the parameters of interest $p(\bm{\phi}_j)$ defined as follows:
\begin{equation}
    p(\bm{\phi}_j) = \sum_d \mathcal{N}P_{\text{NN}}(\bm{\phi}_j|d),
\end{equation}
where $\mathcal{N}$ is a normalization factor which can be computed through marginalization obtaining the following expression:
\begin{equation}
    p(\bm{\phi}_j) = \sum_d P_{\text{NN}}(\bm{\phi}_j|d)\sum_{k=1}^{N_{\phi}}P(d|\bm{\phi}_k)p(\bm{\phi}_k)\delta\phi.
\label{Eq:prior}    
\end{equation}
Here, $P(d|\bm{\phi}_j)$ corresponds to the likelihood function of the system and it governs the system behaviour as a function of the vector of parameters under study. The latter can be approximated with the occurrence frequencies $f_{d,j}$ of each outcome $d$ retrieved from the whole training set. The prior can be then computed solving equation \ref{Eq:prior} as an eigenvalue problem and it is determined from the sampling of the training set (see \cite{nolan2021machine}). 

Once having trained the NN and retrieved both the \emph{a prior} distribution and the single-measurement posterior probabilities, it is possible to perform the estimation applying the Bayes's theorem Eq.\ref{Bayes} to update the prior knowledge depending on the $m$ measurement results:
\begin{equation}
    P(\bm{\phi}_j|\bm{d}) = \bar{p}(\bm{\phi}_j) \prod_{i=1}^m \big(\frac{\bar{P}_{\text{NN}}(\bm{\phi}_j|d_i)}{\bar{p}(\bm{\phi}_j)}\big).
\end{equation}
Here, the upper bar over the probabilities indicates that they have been rescaled for the factor $\delta\phi$. 

\subsubsection*{Extension to adaptive regime}

We have extended such approach to perform the NN Bayesian update in an adaptive framework in the limited data regime. Such protocols are indeed vital for \emph{ab-initio} estimation problems \cite{berni2015ab}, where the possibility to perform the estimate at different working points of the device allows to disambiguate values associated to the same detection probability which therefore results to be a non-monotonic function in the considered parameters interval. Here, the discrimination can be done applying a random feedback after each interaction of the probe with the investigated system. 

In this scenario, the estimate of the parameters of interest is done after each interaction of the probe with the system and a series of control parameters $\bm{c}$ can be tuned after each step of the estimation protocol, setting the system in a different condition in order to increase the amount of information extracted about the parameters. Moreover, to perform the Bayesian update efficiently we use the QInfer implementation \cite{granade2017qinfer} of a particle filtering algorithm also known as sequential Monte Carlo (SMC) \cite{granade2012online} which, in this case, is particularly appropriate since the parameter space is already discretized. 
Indeed, in our implementation the number of points $N_{\phi}$ correspond to the so-called particles of SMC approximation and their initial locations correspond to the grid points in the training set. The integrals are therefore substituted with the respective discrete approximation and the generic probability distribution is replaced by a sum over all the discrete points for the respective weights: $p(\phi)\approx\sum_k w_k\delta(\phi-\phi_k)$. Moreover, in SMC a resampling tecnique is recommended \cite{granade2012online}, which shifts the particle positions to more likely locations during the estimation process, to avoid precision loss due to discretization. However, this last aspect of the technique is not implemented when applying SMC to Bayesian NN, since the latter algorithm is developed for fixed particle positions.

Before sending each probe state, we set the vector of control phases $\bm{c}$ and after each measurement result $d$ the particles weights are updated through the NN Bayesian single-measurement update. However, to assign the right weights, we remove the resampling procedure and we shift both the Bayesian and the prior distribution accordingly, i.e. $P_{\text{NN}}(\bm{\phi}_j-\textbf{c}|d)$ and  $p(\bm{\phi}_j-\textbf{c})$ paying attention to renormalize the updated particles weights:

\begin{equation}
    w_i\rightarrow w_i\Big(\frac{P_{\text{NN}}(\bm{\phi}_j-\textbf{c}|d)}{p(\bm{\phi}_j-\textbf{c})}\Big)\big/ n,
\label{BayesNN}    
\end{equation}
where $n$ is the normalization factor. Note that, with this procedure (Eq.~\eqref{BayesNN}) we generalize to adaptive strategies the protocol of \cite{nolan2021machine}.

\subsection*{Feedback-based Neural Network for single phase estimation}

\begin{figure*}[ht!]
\centering
\includegraphics[width=0.99\textwidth]{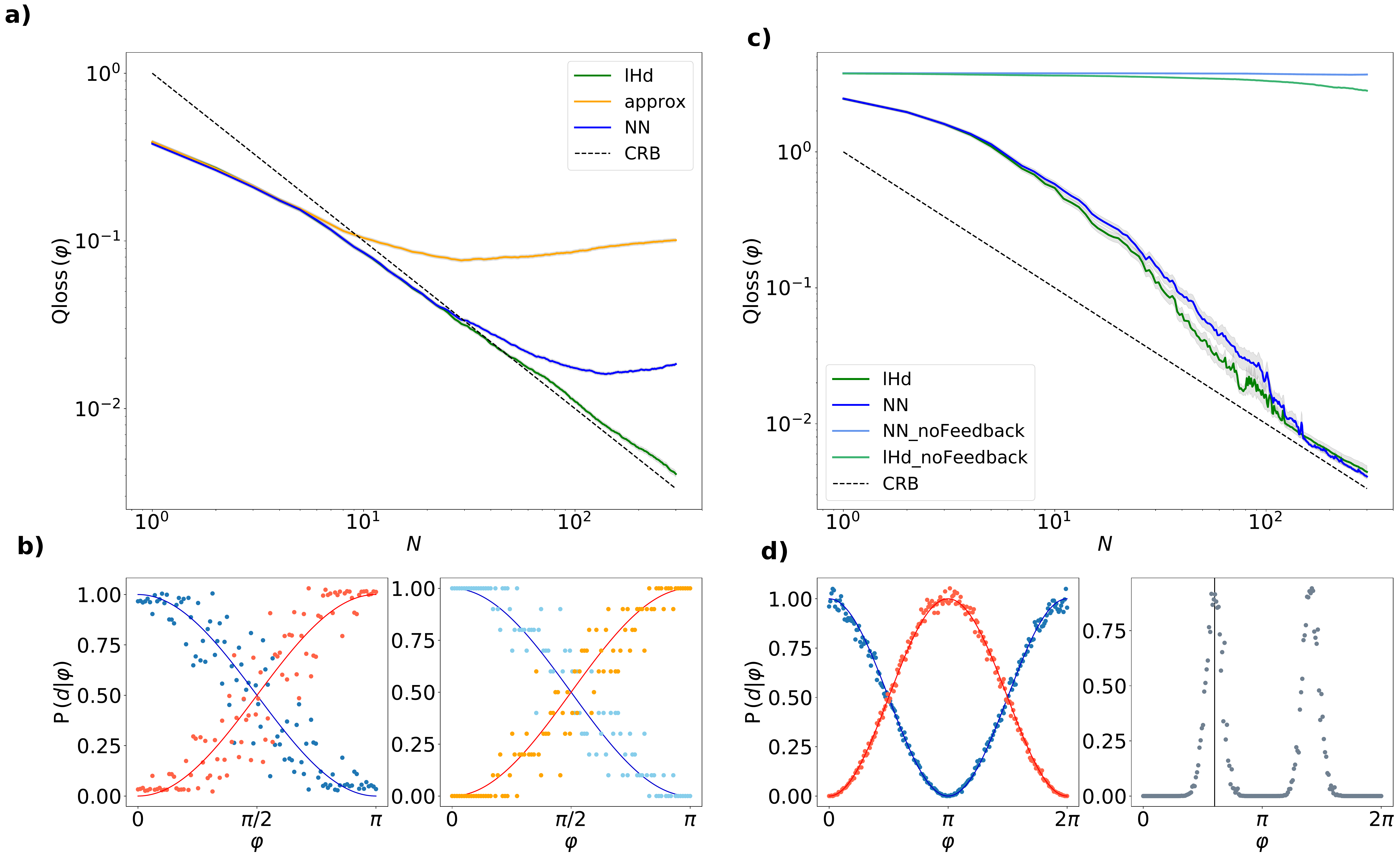}
\caption{Single-phase estimation in a Mach-Zehnder interferometer. \textbf{a)} Averaged quadratic loss, as a function of the number of probes $N$, computed over $30$ repetitions of $100$ phase values of $\varphi\in[0,\pi]$. The results are obtained setting the control phase to zero. We compare the results obtained when having the full knowledge of the outcome probabilities (green line), with the ones achieved using the NN reconstructed single-measurement posterior probability (blu line) and the ones resulting from approximating the likelihood of the system with the occurence frequencies (yellow line) both retrieved performing $r=10$ measurements for each of the $N_\varphi = 100$ grid points. \textbf{b)} Likelihood functions relative to the two possible measurements outcomes reconstructed via the NN on the left and with the standard calibration procedure on the right with $r=10$ and $N_\varphi = 100$ in the $\pi$ interval. The continuous lines represent $P(d|\varphi)$. \textbf{c)} Averaged quadratic loss, as a function of the number of probes $N$, computed over $30$ repetitions of $100$ phase values of $\varphi\in[\epsilon,2\pi-\epsilon]$. Results obtained with the likelihood and the NN update (reported in green and blu respectively) when estimating $\varphi\in[\epsilon,\pi-\epsilon]$ without feedbacks (light green and ligh blu lines) and applying a random feedback after each probe (green and lines). The shaded grey area in the plots represent the interval of one standard deviation while the dashed black line is the SNL$=1/N$. \textbf{d)} Likelihood functions relative to the two possible measurements outcomes reconstructed via the NN obtained for $r=1000$ and $N_\varphi = 200$ in the $2\pi$ interval. On the right is reported the posterior NN probability reconstructed after $20$ probe states were measured. As discussed in the main text, due to the non-monotoncity of the output probabilities in the considered phase interval the posterior shows two peaks and this makes it necessary to use different feedbacks. The black line represents the true value of $\varphi$.} 
\label{fig:MZ}
\end{figure*}

We start applying the designed protocol for the estimation of a phase shift $\varphi$ among two arms of a Mach-Zehnder interferometer.
Bayesian estimation protocols require the knowledge of the likelihood function describing the detection probabilities at the two output ports $d=\{0,1\}$ of the interferometer as a function of the parameter of interest, i.e. $P(0|\varphi) = \cos^2(\varphi/2)$ and $P(1|\varphi) = 1-P(0|\varphi)$.
Probing the system with a sufficient number of probes $N$, it is possible to retrieve an estimate whose performances converge to the ultimate precision bound. Due to the monotonicity of the problem such optimal performances are granted only when $\varphi\in[0,\pi]$, indeed to disambiguate the $\varphi$ values in all the periodicity interval an adaptive scheme must be implemented. Instead of relying on the likelihood knowledge, we train a NN to implement a black-box Bayesian update (see Eq.\ref{BayesNN}). As expected, increasing the number of single-measurement repetitions $r$, corresponding either to the outcome $0$ or $1$, and consequently the training set size, the posterior probability reconstructed during the training becomes more accurate. However, when only a limited amount of measurements is available the estimation precision retrieved through the NN Bayesian update is considerably higher than the one achieved with a standard calibration methods as shown in panel \textbf{a)} of Fig.\ref{fig:MZ}. Here, we compare the estimation performances achieved when the full knowledge of the system is available (lHd), with the ones obtained when performing the Bayesian update through the posterior reconstructed by a NN trained when only $r=10$ measurements for each of the $N_\varphi = 100$ labels of $\varphi$ are available. The performances are compared with a standard calibration procedure approximating the model likelihood with the relative occurrence frequencies extrapolated from the same set of measurement results used for the training (approx). The performances are computed in terms of quadratic loss: $\text{Qloss}(\varphi) = (\sum_iw_i\varphi_i-\varphi_{\text{true}})^2$, to be robust against the presence of possible biases in the estimation procedure, sampling randomly $100$ independent values of $\varphi_{true}\in[0,\pi]$. To make the results more robust we repeat $30$ times the estimation protocol for each inspected phase value. The reported results correspond to the average over all the repetitions and the phase values, the shaded grey area is the region of one standard deviations of such averaged results. The comparison of the achieved performances is done with the shot-noise limit (SNL), i.e. $(\Delta\varphi)^2=1/N$, which corresponds in such a scenario to the ultimate precision bound.
Importantly, the small bias between the bound and the estimation with the likelihood is a consequence of the limited number of particles used to discretize the parameter space which, as previously discussed, is equal to $N_\varphi = 100$. 

Such results show the enhanced performances achieved by the NN Bayesian update compared to standard calibration procedures when a limited set of measurement outcomes is available. To understand the reason for this difference, we show in panel \textbf{b)} of Fig.\ref{fig:MZ} the reconstructed likelihood functions with these two approaches. It can be seen that the NN is able to better disambiguate close values of $\varphi$, while, when simply approximating the probability with the registered occurrence frequencies, close values of $\varphi$ are all associated to the same probability value.
Note that, since the estimation is done by restricting the prior distribution to $[0,\pi)$ domain, we approach the bound from below \cite{d2022experimental} and such results are obtained setting the control phase $c=0$ for all the probe states. As previously stated, setting a value of $c\ne0$, different for each probe, becomes fundamental when $\varphi\in[0,2\pi]$ indeed as shown in panel \textbf{d)} of the same figure the likelihood in this interval is not a monotonic function. The  comparison of the estimation results for a non-adaptive protocol (in light green and light blue) with the ones achieved setting a random feedback after each Bayesian update is reported in panel \textbf{c)} of Fig.\ref{fig:MZ}. As it can be seen the performances achieved with the NN estimation method are close to the ones obtained when the explicit function of the probability outcomes is known. However, in this scenario, the estimation of edge values performs poorly due to the randomness of the selected adaptive feedbacks. Therefore, we choose $\varphi_{true}$ in the reduced interval $[0+\epsilon,2\pi-\epsilon]$, setting $\epsilon=0.3$.

\subsection*{Extension to real scenarios: multiparameter estimation}

\begin{figure}[ht!]
\centering
\includegraphics[width=\columnwidth]{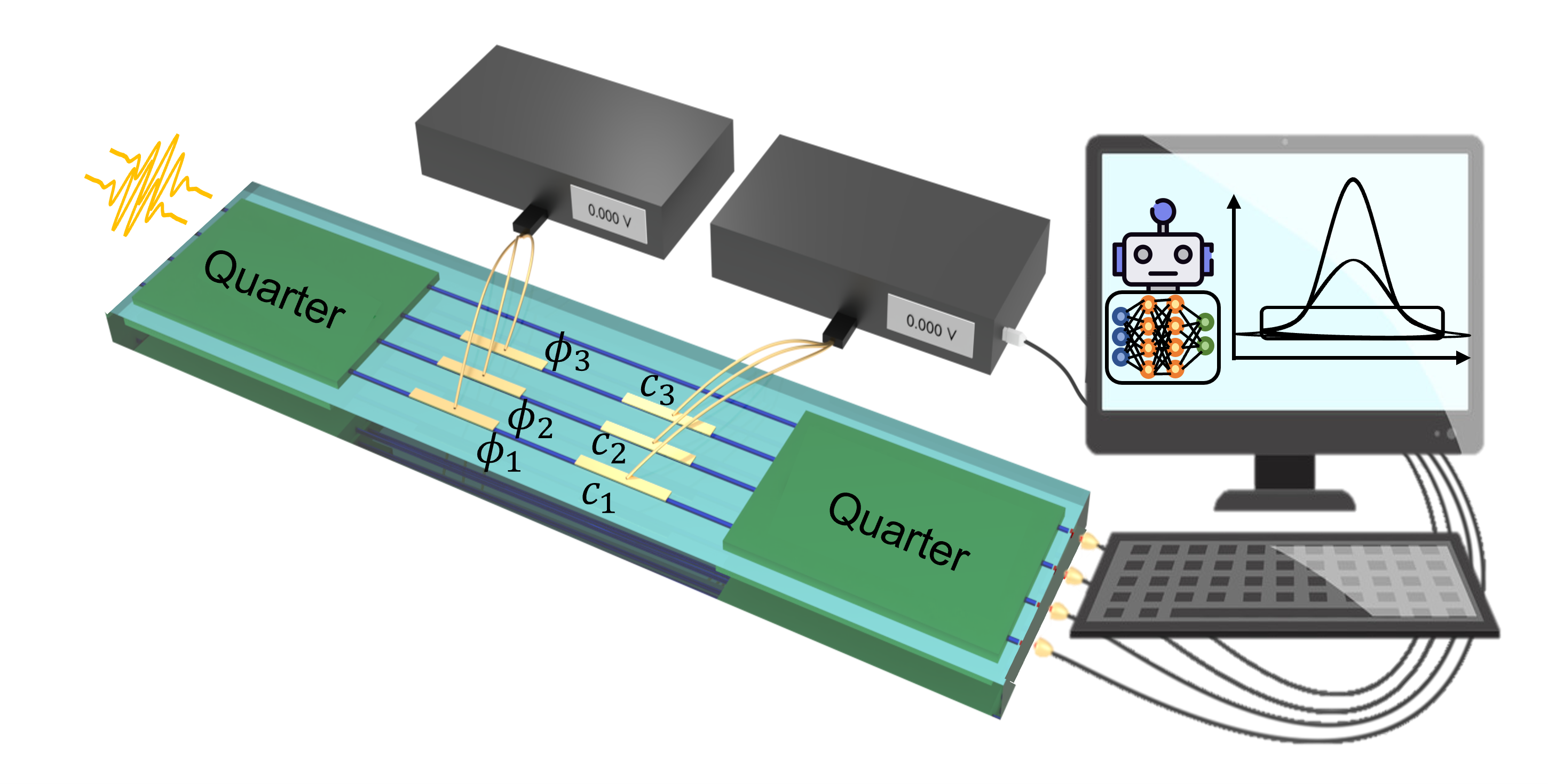}
\caption{Scheme of the integrated photonic phase sensor. The device consists in a four-arm interferometer with the possibility of estimating three optical phases adjusting three relative phase feedbacks through thermo-optic effects. Two-photon states are injected at the device input and both the Bayesian update and the choice of the optimal feedback are done through ML-based protocols trained directly on measurement outcomes.}
\label{Fig:setup}
\end{figure}

The test of the effectiveness of such NN approach on multiparameter sensors becomes fundamental for extending Bayesian estimation protocols to high-dimensional systems where the derivation of the explicit model often requires a great effort and not always is available. Indeed, in a real scenario, the access to any desired quantum state can be particularly challenging or even impossible. Therefore, a calibration of the quantum device, based on the device physical model necessary for its optimal use, is not always feasible and often computationally and experimentally demanding.
We apply the Bayesian NN adaptive protocol to estimate three optical phases in an integrated $4-$arm interferometer \cite{multipara2022}. The device is fabricated through the femtosecond laser writing technique \cite{osellame2012femtosecond} and all the optical phases can be tuned applying a voltage to various microheaters, patterned in a thin gold layer, and placed onto the different arms of the interferometer. In particular, the interferometric phases under study are combined with two integrated quarter splitters (4x4 balanced beam splitters) that close the interferometer (see Fig.\ref{Fig:setup}). The presence of a pair of microheaters in each of the internal arms allows to set both the triplet of investigated phases $\bm{\vec{\phi}}$ and the control feedbacks $\bm{\vec{c}}$ to implement adaptive protocols (see Methods for more details on experimental platform).

In the same spirit of the procedure described for the single-phase Mach-Zehnder interferometer, here we identify each triplet of phases with a specific class in order to train the NN for Bayesian update. We discretize the parameters space building a grid of $N_{\phi}^3$ different triplets. The training is performed associating the single-measurement results to the respective triplet of phases set on the sensor corresponding to an univocal class. 
In order to achieve higher estimation performances we inject into the device pairs of indistinguishable photons, which after the interaction into the first quarter are projected into a two-photon entangled state (see Fig.\ref{Fig:setup}). With the two-photon inputs the possible output configurations are $d=10$: $4$ related to the events having both the photons in the same output port of the integrated device and the $6$ combinations of the two indistinguishable photons in two different outputs of the $4$ ports of the chip. Due to the structure of the output probabilities of our device when two-photon entangled states are injected, we are able to estimate unambiguously phases values in a $\pi$ range. However we need to be able to set feedbacks. From this it follows that the training has to be done in the whole $2\pi$ interval such that $\bm{\phi}+\bm{c}\in[-\pi,\pi]$.

\begin{figure*}[ht!]
\centering
\includegraphics[width=0.99\textwidth]{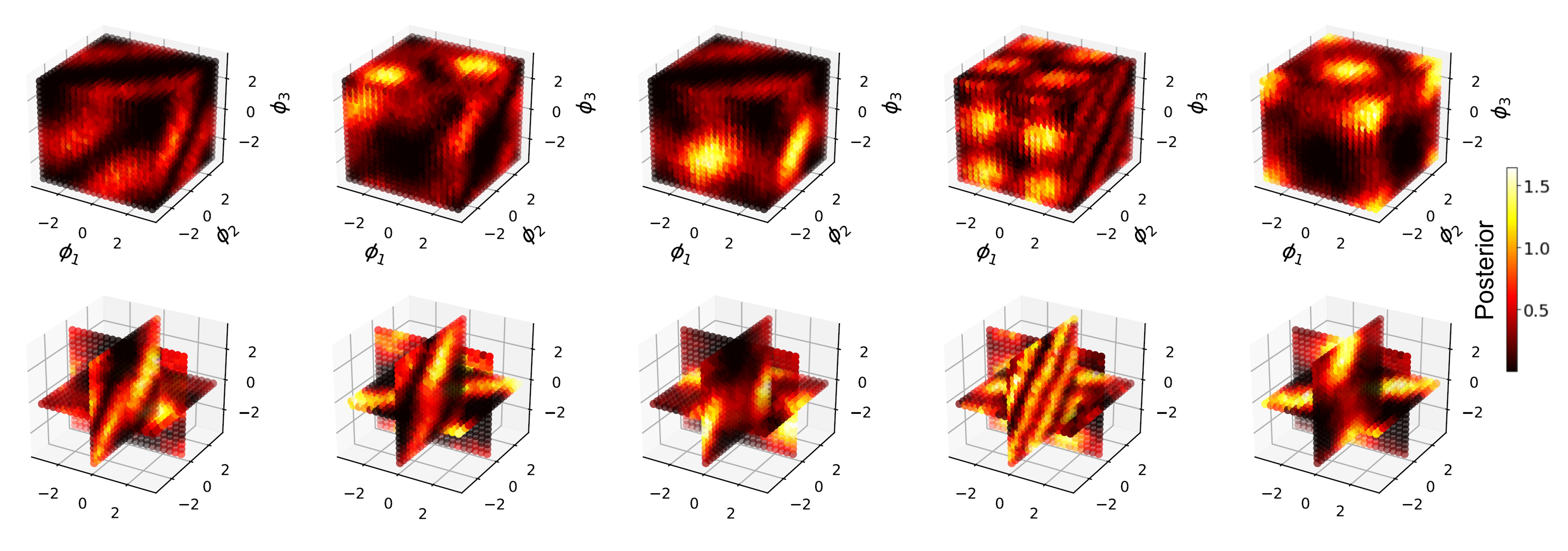}
\caption{Experimental posterior probability distributions reconstructed by the NN. The points on the three axis correspond to the $N_\phi^3 = 8000$ grid points measured while the color indicates the value of the probability. Only half of the $10$ possible probabilities are reported here: in particular, the probabilities relative to $d = 1, 3, 5, 7, 10 $ are shown. In the second row we have reported three slices, of the corresponding above probability, obtained fixing the value of one phase to zero to give more insights on the probabilities structure.}
\label{fig:PNN}
\end{figure*}

To ensure the achievement of the optimal estimation performances a sufficient number of grid points $N_{\phi}$ and of measurement repetitions $r$ is required. To identify the minimum size of the training set we perform some simulations with different grid spacing changing $N_{\phi}$ and different number of measurement repetitions $r$ (see Supplementary). All the simulations are done using the likelihood function of the ideal device to simulate the measurement outcomes. We therefore choose $N_{\phi}$ and $r$ allowing both to achieve good performances, collect the necessary data, and perform the training in a reasonable time. To satisfy these conditions, we set $N_{\phi}=20$ corresponding to $20^3$ different triplets of phases in the interval $[-\pi,\pi]$ and collecting for each one $r=1000$ events. 
The training is performed directly on experimental data corresponding to the measurement of one of the $10$ possible output configurations each associated to the corresponding vector of the $N_\phi$ parameter labels. Details on the network architecture are reported in the Methods. 
Notably, the extension to the multiparemeter scenario has required additional computational efforts related to the huge dimension of the training matrices (see Methods). 
Once trained, we can reconstruct the posterior probability associated to each of the $10$ measurement outcomes for all the grid points. The obtained results for half of the probabilities are reported in Fig.\ref{fig:PNN} (the other five probabilities are plotted in the Supplementary).

We start by inspecting the performances achieved applying random feedbacks after each probe and then we implement an optimization algorithm through RL to select the feedbacks assuring a faster convergence to the bound, a fundamental requirement in the limited data regime.

Once the NN for the single-measurement Bayesian update is trained, we implement an estimation protocol which uses the posterior probability learned by the network directly from the experimental data to update the knowledge on new experiments. Since the prior distribution is determined by the training data, we have to rescale all the probabilities derived by the NN training to solve the monotonicity issues of our estimation problem. We start shifting the NN probabilities $P_{\text{NN}}(\bm{\phi}_j-\textbf{c}|d)$ and  $p(\bm{\phi}_j-\textbf{c})$, as seen before, in the whole periodicity interval to take into account the value of the feedbacks, but before performing the Bayesian update we select only the values in the $\pi$ interval $\bar{P}_{\text{NN}}$ and $\bar{p}$, renormalizing the obtained probabilities as follows:
\begin{equation}
\begin{split}
P_{\text{NN}}^{new} = \bar{P}_{\text{NN}} \Big/ \Big(&\sum \bar{P}_{\text{NN}} \cdot \frac{\pi}{(\nicefrac{N_{\phi}}{2})^3-1} \Big),\\
p^{new} = \bar{p} \Big/ \Big(&\sum \bar{p} \cdot \frac{\pi}{(\nicefrac{N_{\phi}}{2})^3-1} \Big).
\end{split}
\end{equation}

We perform the protocol offline. First of all we collect the events relative to a grid of phases with more statistics than the one used for the NN training, allowing us to compute the outcomes probabilities associated to all the grid points. Then, we select a random triplet of phases in the prior and the events at each step of the estimation protocol are picked from the experimental grid with the relative probability.

\subsection*{Reinforcement learning for black-box adaptive quantum metrology}

The possibility to implement adaptive protocols becomes fundamental, in the limited data regime, to speed up the estimation process by adopting a practical measurement scheme based on feedbacks coming from the measured system \cite{wiseman1995adaptive}. The use of feedbacks allows to optimize the protocol when only a limited number of probe states can be used for the estimation.

We now combine the demonstrated NN Bayesian update with a new concomitant learning agent interacting with the NN output.
More specifically, we implement a RL algorithm that, using the NN update, sets the optimal control parameters to ensure a faster convergence 
of the estimation
with the minimum amount of resources. For high-dimensional and complex systems the convergence to the ultimate precision bound, with a limited number of probes, is indeed granted only if at each step of the protocol the relative optimal feedbacks are set. This allows to extract the maximum amount of information from each probe state. 

\subsubsection*{RL-based design heuristics}

The purpose of RL is to find an optimal strategy, often referred to as policy, that the agent can perform on the environment in order to maximize its reward. In particular, the policy represents the conditional probability distribution $\pi(a|s)$ of performing the action $a$ conditioned on the observed environment state $s$.
For problems with continuous action spaces, the agent’s policy can be modeled as a parameterized function of states such as deep neural networks.
The method that we chose for the RL algorithm is the cross entropy method (CEM) which is one of the most generic and easy to implement methods. It maximizes the agent's reward with a derivative free optimization approach, it can be considered as a black-box approach since it looks for the NN weights linked to actions gaining the highest reward. 
Such weights $\bm{\omega_i}$ are initially sampled from a Gaussian distribution with a given mean and variance: $\bm{\omega_i}\sim\mathcal{N}(\mu,\sigma)$. Then, $n$ batches of episodes are sampled from the distribution in which the agent performs some actions from the policy network based on the relative weights and the rewards generated by the environment for each episode are registered.
Every episode consists on a sequence of observations of states of the environment when the agent makes actions.
Only episodes showing a reward above a certain threshold are kept and such \emph{elite} weights of the relative NNs are used to compute a new mean $\mu$ and variance $\sigma$ for the new weights distribution. For this reason such method is also called an evolutionary algorithm since it samples the NN weights from a distribution which is updated at each iteration. Such procedure is iterated until the mean average reward for the batch of episodes converges to the desired value. 

\begin{figure}[h!]
\centering
\includegraphics[width=0.99\columnwidth]{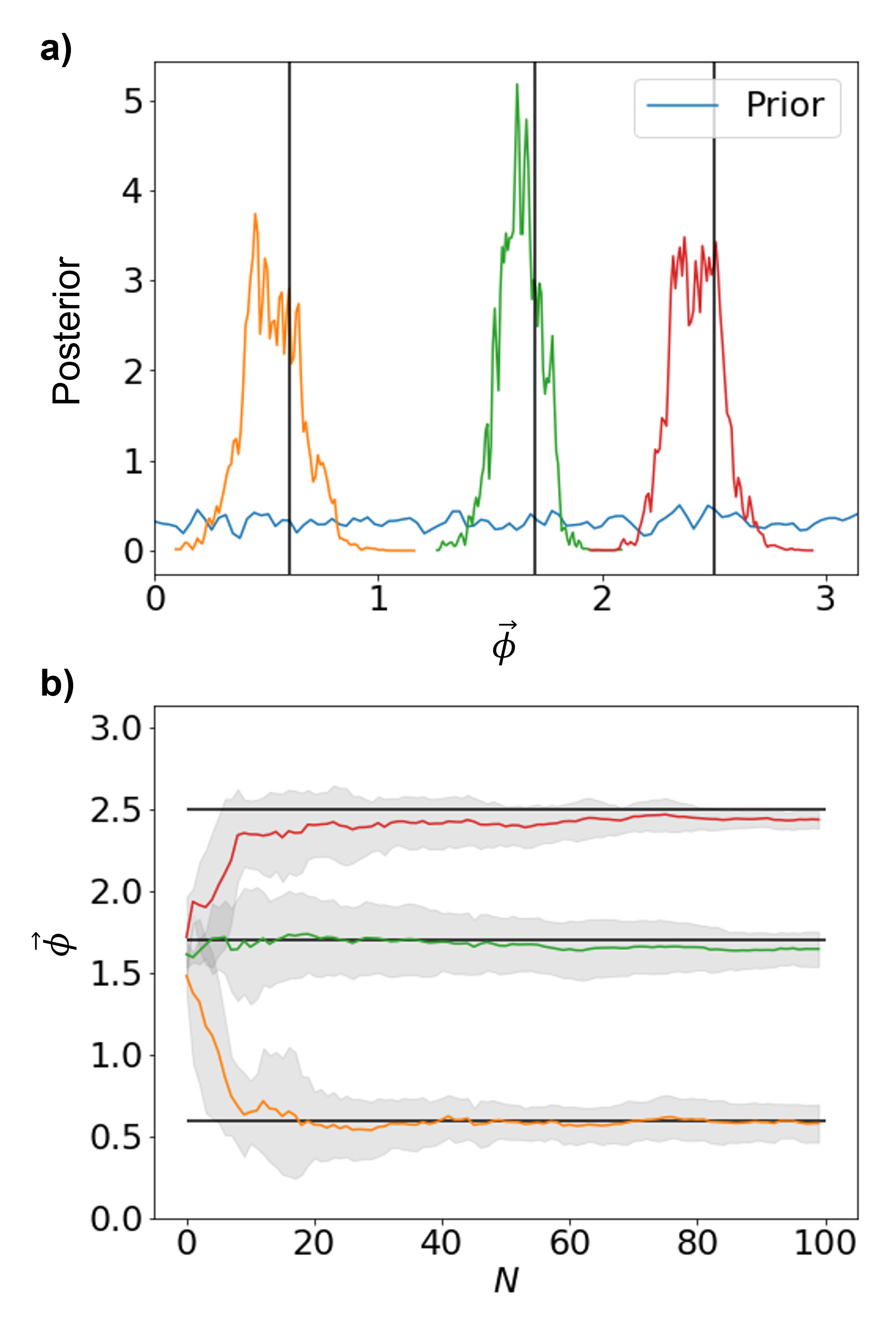}
\caption{Estimate of $\bm{\phi} = [0.6, 1.7, 2.5]$rad retrieved applying the standard Bayesian estimation using the likelihood of the ideal device and optimizing the control feedbacks with the RL agent. \textbf{a)} The blue line represents the prior distribution, while the orange, green and red lines are the reconstructed posterior probabilities for the first, second and third phase respectively. \textbf{b)} Estimated values as a function of the number of probes. Continuous lines represent the average over $30$ repetitions while the grey area is the interval of one standard deviation.}
\label{fig:stima}
\end{figure}


The training is performed offline using the same grid of data used for the training of the Bayesian NN. At each episode, a true value of $\bm{\phi}$ is sampled from the prior distribution and the agent performs a sequence of actions, depending on the number $N$ of available resources, setting the control phases $\bm{c}$, therefore, changing the operation point of the device. The obtained measurement outcomes are selected from the grid point closer to the imparted phase shift. After each measurement outcome, it is possible to update the posterior probability distribution with the one retrieved by the Bayesian NN. When such sequence is finished, it is possible to compute the reward function achieved with these settings. We choose as reward function the one used in \cite{fiderer2021neural}:

\begin{equation}
    R(d_m) = \text{Tr}[\text{cov}_{\bm{\phi}|d_{m-1}}(\bm{\phi})]-\text{Tr}[\text{cov}_{\bm{\phi}|d_{m}}(\bm{\phi})],
\end{equation}
which is the difference in the traced covariance over the posterior distribution after the updating with the measurement result obtained with a new probe.
When all the $N$ probes are used the episode is completed and the environment is reset, the posterior is reset to the prior and a new training episode starts.

It is important to note that, differently from Fiderer et al. \cite{fiderer2021neural}, using the NN Bayesian update we do not rely in any way on the system model. This crucial step, combining the aforementioned techniques, allows us to reach a totally black-box approach for metrological tasks.



\begin{figure*}[ht!]
\centering
\includegraphics[width=\textwidth]{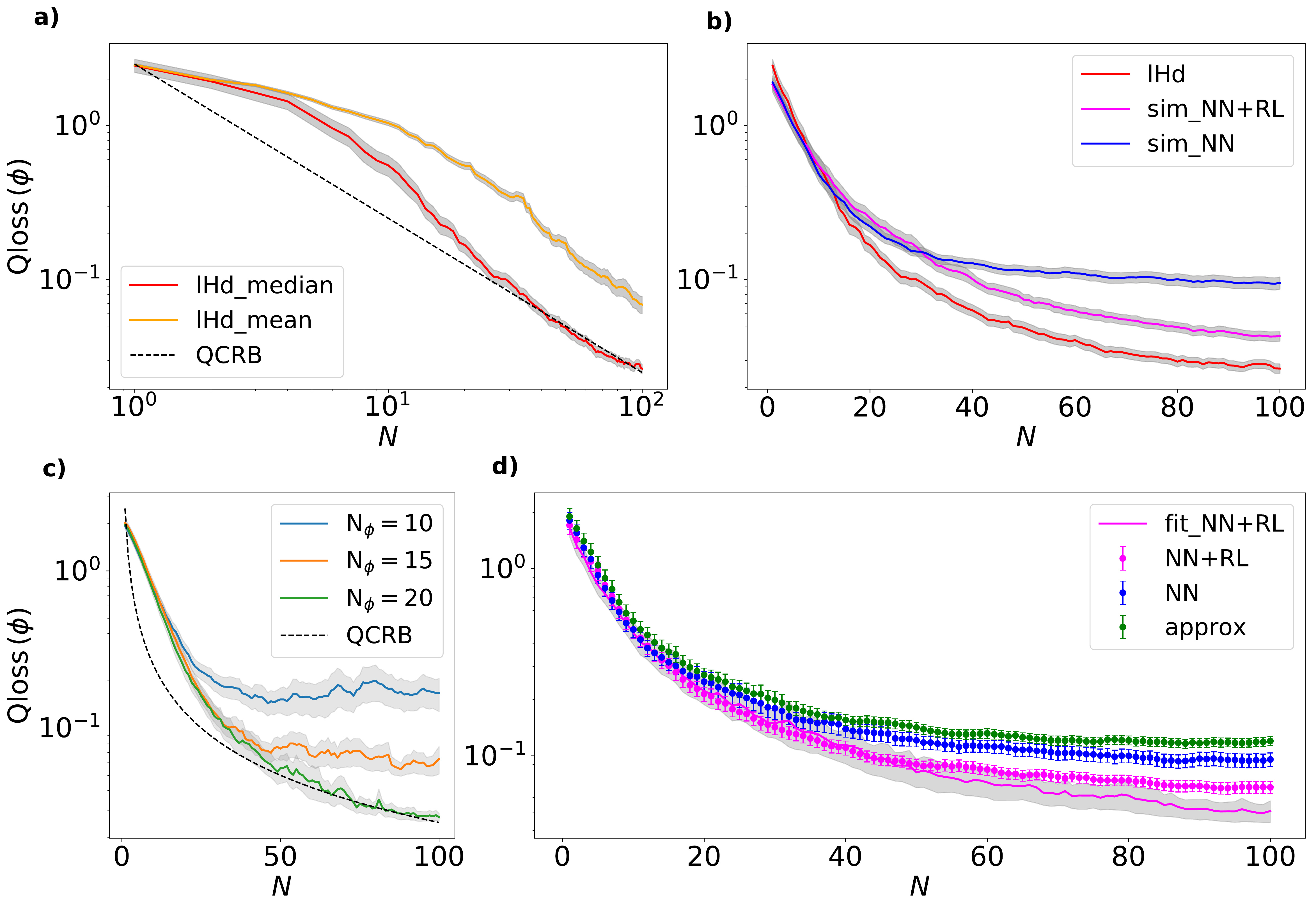}
\caption{Three-phase estimation in a 4-arm interferometer. Achieved Qlosses averaged over $100$ different triplet of phases in the interval $(0,\pi]$ as a function of the number of probes. The grey area represent the standard deviation on the mean values. \textbf{a)} Performances of the ideal device obtained when the explicit model is used for the Bayesian estimation. The orange line represents the mean over all the $30$ repetitions for each of the $100$ parameters inspected while the red line is the median over the different repetitions. The dashed line is the QCRB that for our device is $2.5/N$. \textbf{b)} Average over $100$ triplet of phases of the median Qloss computed over $30$ repetitions of the estimation protocol. Comparison with the results obtained when substituting the Bayesian updated through the explicit posterior (red line) with the one reconstructed by a NN trained on simulated data (magenta line). The blue line represents instead the performances achieved applying random feedbacks instead of the ones found by the RL agent. \textbf{c)} Simulation on the ideal device changing the number of grid points $N_\phi$ in the training of the Bayesian NN. Since the training for such simulations has been done in the restricted interval $[0,\pi]$ here we limit the possible applied feedbacks to satisfy the condition $\bm{\phi}_{true}+\bm{c}\in(0,\pi]$. \textbf{d)} Experimental results achieved with the Bayesian NN update and the RL optimization algorithm (magenta points), when the latter is substituted by a random choice of feedbacks (blu points) and when the Bayesian update is done approximating the likelihood with the occurrence frequencies (green points). Error bars represent the standard deviation of the averaged Qlosses. The magenta line shows the performances obtained with simulation done using the likelihood function of the real device and it is shown as a reference.  
}
\label{fig:final}
\end{figure*}

\subsubsection*{Experimental results}

As a benchmark model, we use the RL protocol to optimize the feedbacks in a simulated experiment of an ideal $4-$arm interferometer. In this case, we can compute the likelihood of the ideal device using it both to reconstruct the posterior distribution necessary at each step of the Bayesian estimation protocol and to simulate the measurement results. We demonstrate that the trained agent allows to select the optimal feedbacks necessary to show a faster convergence for a smaller number of probe states $N$ than the one obtained if setting random controls. We train the agent on $10^4$ episodes, simulating the measurement outcomes obtained after the choice done by the agent on the control phases. 
We perform the simulation with the ideal likelihood function using the same number of particles that will be used in the NN based approach i.e. $\text{n}_\text{PART} = N_\phi^3 = 10^3$. In panel \textbf{a)} of Fig.\ref{fig:stima} we show the prior distribution and the reconstructed posterior after sending $N=100$ probe states for a specific triplet of phases. The averaged estimate on $30$ different repetitions of the experiment on the same triplet are reported, with the corresponding standard deviation, in panel \textbf{b)} of the same figure, as a function of the number of sent probes. 

The results achieved with the RL optimization in terms of Qloss, when the explicit model of the system is known, are reported in panel \textbf{a)} of Fig.\ref{fig:final}. The dashed line represents the ultimate precision bound corresponding to the Quantum Cramér-Rao bound (QCRB) \cite{paris2009quantum} of the ideal device injected with the employed input states, while the red and the orange lines represent the averaged performances over $100$ different triplet of phases after performing the median and the mean respectively over $30$ different repetitions of the Bayesian SMC protocol. Note that, the distribution of errors in phase estimation of such protocol indeed shows fat tails due to the presence of outlier experiments where the protocol fails as already noticed in \cite{wiebe2016efficient}. In order to reduce the weight of such outliers, the median instead of the mean is commonly used as figure of merit since, the former is less sensitive to the presence of few outliers. 

We then compare the obtained results with the full knowledge of the system with the ones achieved when the Bayesian update is done through the single-measurement posterior reconstructed by the NN. In this case, also the training of the RL algorithm is done updating the posterior after each choice of the agent through the NN reconstructed probability distribution. We therefore generate a grid of simulated data using the likelihood function of the ideal device, for the Bayesian NN training, with the same step size and configurations used before on the experimental data. In panel \textbf{b)} of Fig.\ref{fig:final} we can see that the performances of the optimization algorithm are still very high, the remaining difference among the red curve, obtained when knowing the likelihood of the system, and the magenta curve is related to the fact that to train the Bayesian NN we choose a grid of only $N_\phi = 10$ points in the interval $[0,\pi]$ for each of the three phases (see Supplementary for further investigations). The improvement obtained performing the optimization of the control phases is clearly visible when comparing the results obtained with the same network but randomly choosing the set of feedbacks after each measurement result (blue curve). Only with the aim of studying the impact of the reduced size of the grid, we study the performances achieved when training the NN in the reduced interval $[0,\pi]$. Consequently, only for this test, we need to restrict the possible random feedbacks in the interval $[-\phi_{\text{true}},\pi-\phi_{\text{true}}]$. The results obtained for 
different discretizations $N_\phi$ are reported in panel \textbf{c)}. Even if the estimation results  are not independent from the true value $\phi_{\text{true}}$, this is a clear proof that using grids with a higher number of points assures optimal performances also when using the NN-based Bayesian update \footnote{Note that, it is possible to reach the saturation in this scenario with random feedbacks, only because of the implicit knowledge on the true parameter values when setting such feedbacks.}.

Finally, we test such algorithm on the actual device reporting in panel \textbf{d)} of Fig.\ref{fig:final} the experimental results. Here, the training of both the NN for Bayesian update and the one of the RL agent are done directly on the grid of collected experimental data. The reported Qlosses refer to the average over $100$ triplet of phases of the median Qloss computed over $30$ repetitions of the estimation protocol of different experiments.
For the sake of demonstrating the goodness of the experimental results, we perform some simulations using the likelihood function of the real device reconstructed from experimental grids knowing the model of the system (this is equivalent to have infinite resources for the system calibration). 
As expected the performances on the actual experimental data are slightly lower than the ones achieved with the simulated data (magenta line) due to the presence of experimental noise into the training data. However, even for the protocol trained directly with experimental data, the improvement of the multiparameter estimation precision when using the RL algorithm combined with the NN Bayesian update is clearly visible compared to the performances obtained, with equal resources, by methods not using the optimal feedbacks or an approximated likelihood without the NN learning. 
In this way, the results in  Fig.\ref{fig:final} demonstrate the advantage of a full artificial intelligence approach for black-box quantum metrology.



\section*{Conclusions}
Quantum sensing represents one of the most promising applications of quantum theory. In order to develop optimized quantum metrology protocols, one has to face several challenges in the limited resources regime: the characterization of the quantum sensor operations as well as devising optimal feedback for adaptive estimations.
In this work, we overcome these fundamental challenges by developing a deep reinforcement learning protocol which combines a RL agent, designated to choose the optimal control feedbacks, with a deep NN which updates the knowledge on the parameter values, in an actual noisy multiparameter estimation experiment. The quantum sensor is represented by an integrated photonic circuit consisting of a 4-arm interferometer, seeded by indistinguishable photons, for sensing of multiple optical phases. All the NN trainings are performed directly on experimental data, whithout any a-priori knowledge of the considered quantum sensor and relying only on the accessible output statistics of the limited number of set phases points. To achieve these results, we started by generalizing the NN Bayesian updater \cite{nolan2021machine} to the multiparameter case and further extending the protocol for adaptive implementations. Then, we managed to implement such black-box ML approach for the learning of optimal feedbacks. An additional ML protocol, consisting of a RL agent who takes as input the results of the NN Bayesian update is implemented. 
The fusion of these two extended ML algorithms enabled to experimentally demonstrate a fully artificial intelligence approach outperforming standard techniques for optical sensing.

We can implement such automated protocol thanks to the use of a programmable integrated photonic circuit which allows to control the performed measurements, easily configuring control parameters to implement adaptive protocols in a fully black-box fashion using quantum states.

The implementation of a model-free approach for quantum multiparameter estimation paves the way for everyday automated use of complex quantum sensors without the need of time and resource-consuming characterization or the requirement of a faithful theoretical modeling. The latter indeed can be a fundamental limitation in all the scenarios where the theoretical description of the whole quantum evolution is lacking. Consequently, most of quantum metrology scenarios, ranging from microscopy and imaging to Hamiltonian learning will largely benefit from the developed strategy.

\section*{Ackwnowledgments}
This work is supported by the ERC Advanced grant QU-BOSS (Grant Agreement No. 884676), by Ministero dell'Istruzione dell'Universit\`a e della Ricerca (Ministry of Education, University and Research) program ``Dipartimento di Eccellenza'' (CUP:B81I18001170001).

\section*{Materials and Methods}

\subsection*{Experimental details}

The integrated device realizing the quantum sensor is a $3.6$ cm long tunable 4-arm interferometer seeded by pairs of indistinguishable single photons. The chip is realized by means of the Femtosecond Laser Writing (FLW) technique \cite{corrielli2021femtosecond, meany2015laser} able to write waveguides inside a glass substrate, suitable for photons at $785$ nm. 
More specifically, the interferometer is realized by two cascaded 4-arm beam splitters (quarters), each composed of four couplers in a three-dimensional configuration \cite{meany2012non}, whose global action on incoming photons is to equally split the photonic amplitude among the four output modes. 
The four output modes of the first quarter are connected to the inputs of the second quarter through four straight waveguides equipped with thermo-optic phase shifters. These allow to tune the internal phase shifts $\phi_i$ between the arms of the interferometer by applying a current on resistors, with a dissipated power responsible of the the phases changes  \cite{flamini2015thermally,ceccarelli2020low}. The relation linking the dissipated power to the value of the phase shift is approximatively quadratic \cite{Valeri2020}. The three internal phase shifts $\phi_{1}$,$\phi_{2}$ and $\phi_{3}$, with respect to a reference one, represent the unknown parameters to be simultaneously estimated in our sensing problem (see Fig.\ref{Fig:setup}). While the first quarter allows to prepare the photonic probe, the final quarter acts as a measurement operator together with the single photon detectors at the output modes of the interferometer. In order to detect events where the two photons exit along the same output port a probabilistic photon-number resolving detection is employed by means of fiber-beamsplitteters before the detectors. The input and output modes of the chip are pigtailed with single mode fibers. 

The chip is provided by overall $12$ thermo-optic phase shifters.
Two pairs of resistors are devoted to tune the action of the two quarters, respectively. The remaining $8$ resistors are used to set the internal phases. In particular, $3$ resistors are used to set the phases to be estimated, while other $3$ are used to tune the control feedbacks chosen by the RL agent to optimize the estimation process.  

The input used to seed the circuit is composed by two indistinguishable photons injected along the last two input modes, that in the Fock basis results $|0011\rangle$. The probe state before interacting with the unknown phases to be estimated is generated by the input two-photon state evolved by the first quarter. The generated state is a generalized N00N-like state in the $4$-dimensional Hilbert space of the evolution. Given the generated probe state, in the case of ideal quarter, we can calculate the corresponding Quantum Fisher Information (QFI) matrix associated  to the ultimate quantum precision bounds achievable by any estimation procedure for the interferometer phases \cite{polino2020photonic}. More specifically, from the QFI matrix, the bound on the sum of the errors of the three independent phase shifts is equal to $2.5 /N$, where $N$ is the number of two-photon probes employed in the measurement. In Fig.\ref{fig:final} we use this bound as a comparison for the estimation performances achieved by our artificial intelligence protocol.

Note that, in order to obtain the maximum sensitivity with such probe states, the two photons have to be indistinguishable. To guarantee the temporal indistinguishability of the photons inside the chip, one of the photon passes along an optical delay line, able to tune the temporal delay with respect to the other.

The two photons are generated at $785$nm by a degenerate SPDC source composed of a pulsed laser at $392.5$nm.

\subsection*{Bayesian neural network}

The NN architecture used for Bayesian update on the experimental data is a $5$ layers, full-connected network implemented using the python library \emph{keras}. They consist in an input one-node layer and three $64$-nodes hidden layers. The output layer has instead a number of nodes which depends on the discrete grid of acquired data, i.e. $N_\phi^3$ nodes, for our three-parameter problem. All the nodes, except for the output ones, which are activated by a softmax function, are activated by a rectified linear unit (reLu) function initializing their weights with random values extracted from a normal distribution centered at zero and with variance $\sigma^2 = 2/n$, where $n$ is the number of neurons in the previous layer. 
To speed up the training process, the whole training set is divided into $256$ small random batches, which are iteratively analyzed during each training epoch. We train the algorithm for $60$ epochs using as loss function the categorical cross-entropy and the ADAM optimization algorithm \cite{Adam}.

Concerning the training set, for each class a fixed number of measurement repetition $r$ has to be shown to the NN during the training, allowing it to learn the correct conditional probability distribution $P_{\text{NN}}(\bm{\phi}_j|d)$ of the measurement outcomes. Therefore, in the multiparameter scenario the whole training set consist of a one-dimensional input vector $X$ containing in each of its rows the measurement outcomes $d$ obtained for $N_{\phi}^3 \cdot r$ different measurements and an output classification vector $Y$ with the same number of rows and $N_{\phi}^3$ columns. Thus additional computational efforts related to the huge dimension of the training matrices has been required. To solve such issues since we deal with sparse matrices where most of the matrices elements are zero values, we work with the corresponding matrix of coordinates in order to keep only the information of the non-zero values and their position in the high-dimensional matrix.

\subsection*{Reinforcement lerning algorithm}

The training of the RL agent is performed through the cross-entropy method (CEM), following the STABLE-BASELINES \cite{stable-baselines} implementation.
For each iteration of the algorithm the agent picks 
an action from the policy network whose weights are
selected through the CEM. The choice is payed with a reward depending on the vector of observations extracted on the environment. Such vector consists of 
the estimated value $\hat{\phi}$, the current number of adopted probes and the posterior covariance matrix.
Given such observations, depending on the goodness of the implemented action, the reward is defined.

The structure of the network used to train the agent has an input layer with a number of nodes equal to the length of the observation vector, a $16-$node hidden layer and an output layer with three nodes: one for each control feedback.
The hidden layer is activated via a reLu function while the output layer activation is a sigmoid function.


\bibliographystyle{apsrev4-1}
\bibliography{biblio}

\end{document}


\title{Supplementary Information for Deep reinforcement learning for quantum multiparameter estimation}

\author{Valeria Cimini}
\affiliation{Dipartimento di Fisica, Sapienza Universit\`{a} di Roma, Piazzale Aldo Moro 5, I-00185 Roma, Italy}

\author{Mauro Valeri}
\affiliation{Dipartimento di Fisica, Sapienza Universit\`{a} di Roma, Piazzale Aldo Moro 5, I-00185 Roma, Italy}

\author{Emanuele Polino}
\affiliation{Dipartimento di Fisica, Sapienza Universit\`{a} di Roma, Piazzale Aldo Moro 5, I-00185 Roma, Italy}

\author{Simone Piacentini}
\affiliation{Dipartimento di Fisica, Politecnico di Milano, Piazza Leonardo da Vinci, 32, I-20133 Milano, Italy}
\affiliation{Istituto di Fotonica e Nanotecnologie, Consiglio Nazionale delle Ricerche (IFN-CNR), Piazza Leonardo da Vinci, 32, I-20133 Milano, Italy}

\author{Francesco Ceccarelli}
\affiliation{Istituto di Fotonica e Nanotecnologie, Consiglio Nazionale delle Ricerche (IFN-CNR), Piazza Leonardo da Vinci, 32, I-20133 Milano, Italy}

\author{Giacomo Corrielli}
\affiliation{Istituto di Fotonica e Nanotecnologie, Consiglio Nazionale delle Ricerche (IFN-CNR), Piazza Leonardo da Vinci, 32, I-20133 Milano, Italy}

\author{Nicol\`o Spagnolo}
\affiliation{Dipartimento di Fisica, Sapienza Universit\`{a} di Roma, Piazzale Aldo Moro 5, I-00185 Roma, Italy}

\author{Roberto Osellame}
\affiliation{Istituto di Fotonica e Nanotecnologie, Consiglio Nazionale delle Ricerche (IFN-CNR), Piazza Leonardo da Vinci, 32, I-20133 Milano, Italy}

\author{Fabio Sciarrino}
\email{fabio.sciarrino@uniroma1.it}
\affiliation{Dipartimento di Fisica, Sapienza Universit\`{a} di Roma, Piazzale Aldo Moro 5, I-00185 Roma, Italy}

\maketitle

\section{Effects related to the finite-size of the Bayesian neural network training set.}

The reconstruction of the single-measurement posterior distribution done by the neural network (NN) for Bayesian update will depend on the number of points in the training grid. Given the complexity of our multiparameter problem we inspect the results achieved with $N_\phi = {10,15,20}$, corresponding to a total of ${1000,3375,8000}$ grid points respectively in the interval $[0,\pi]$. 
The overall performances will depend also on the number of repetitions $r$ of the single-measurement results per grid points present in the training set. The dependence on both $N_\phi$ and $r$ of the performances achieved in terms of Qloss on simulated data are reported in Supplementary Fig.\ref{fig:1}.
The total number of grid points will also determine, as discussed in the main text, the number of particles used in the Bayesian protocol. For what concerns the experiment, in order to implement the adaptive protocol we choose to collect a grid of $8000$ points in the extended region of $[-\pi,\pi]$, to have the possibility of applying feedbacks in the interval $[-\pi,0]$. Such a choice is at the origin of the difference among the results achieved with the Bayesian update done through the explicit system's model (red line) and the one computed via the NN probabilities (magenta line) reported in panel b) of Fig.6 of the main text.

\begin{figure*}[ht!]
\centering
\includegraphics[width=\textwidth]{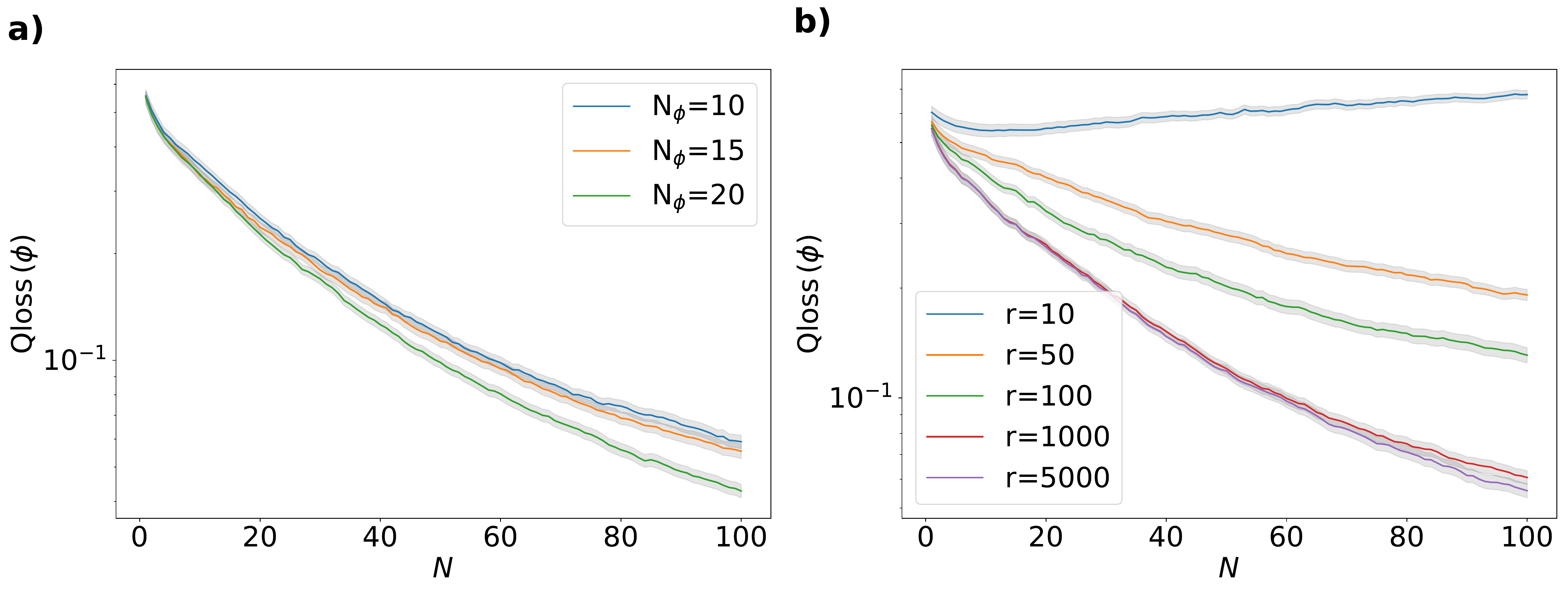}
\caption{Qloss as a function of the number of probes $N$ interacting with the investigated system. \textbf{a)} Results achieved with training set of different size. \textbf{b)} Results obtained with different measurements statistic.}
\label{fig:1}
\end{figure*}

In this scenario the prior and consequently the single-measurement posterior distribution learned by the NN are defined in the $2\pi$ interval and are reported in Supplementary Fig.\ref{fig:priors} and Fig\ref{fig:PNNs}. However, for the estimation procedure we have to consider only the $[0,\pi]$ interval. Therefore, at each step of the protocol we first shift the reconstructed \emph{a-priori} and posterior distributions depending on the set values of the control phases and then we cut the shifted distributions in the interval of interest. The Bayesian update will be done with the reduced distribution using therefore $8000/2^3$ particles.

\begin{figure*}[ht!]
\centering
\includegraphics[width=\textwidth]{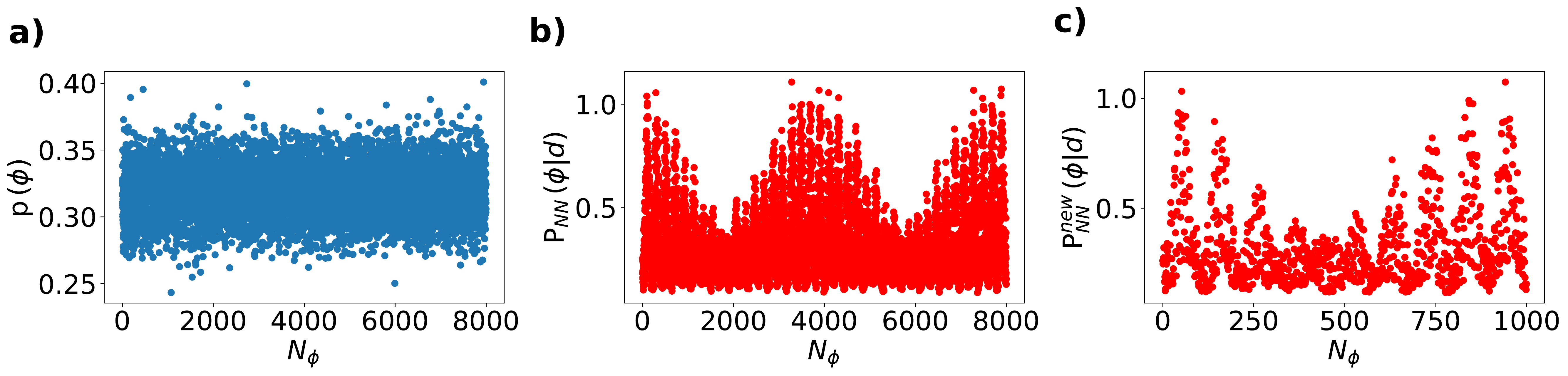}
\caption{\textbf{a)} Prior distribution reconstructed for each data point of the training set.\textbf{b)} Bayesian posterior NN distribution for the outcome $d=6$. \textbf{c)} Posterior probability used for the Bayesian update, relative to the outcome $d=6$, in the reduced interval $\bm{\phi}\in[0,\pi]$.}
\label{fig:priors}
\end{figure*}

\begin{figure*}[ht!]
\centering
\includegraphics[width=\textwidth]{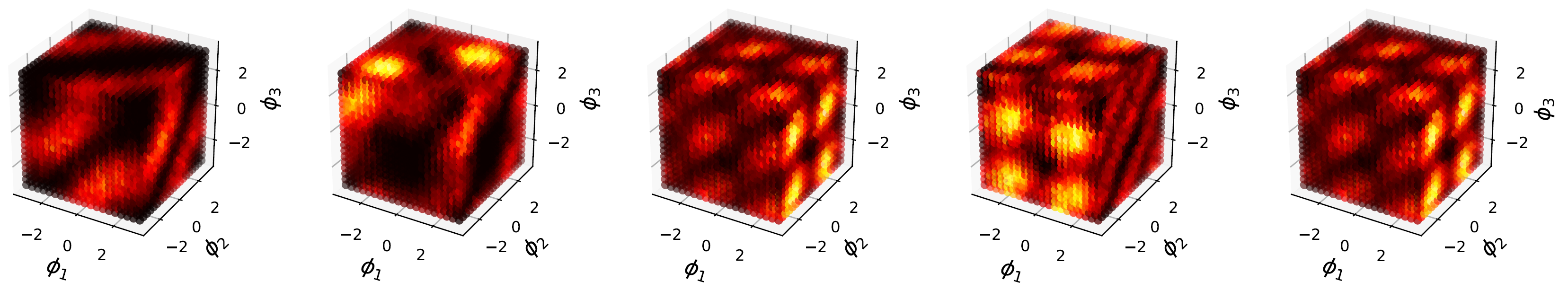}
\caption{Experimental posterior probability distributions reconstructed by the NN for the outcomes $d = 2, 4, 6, 8, 9$.}
\label{fig:PNNs}
\end{figure*}

\section{Training of the reinforcement learning algorithm}

The training of the RL agent has been performed through the cross-entropy method analyzing the results of $10^4$ different episodes for each of the $100$ available probes. The initial weights of the $100$ NN selecting the action performed by the agent are sampled from a gaussian distribution initially centered in zero, with standard deviation $\sigma=0.5$ and fixing the threshold selecting the \emph{elite} weights to the $10\%$ of the overall sample. Once trained we use the NN with the selected \emph{elite} weights to compute the Bayes risk on $100$ independent experiments all consisting of $100$ different measurements. The results obtained when the RL algorithm is run using the likelihood function for both the Bayesian update and for simulating the measurement outcomes are reported in blue in panel a) of Supplementary Fig.\ref{fig:BayesRisk}. Here, the Bayes risk is compared with the one achieved when substituting the explicit posterior with the NN probability learned on the grid of measured data and the experimental results are selected with the respective probability from the same grid (orange line). Due to the presence of experimental noise and to the difference among the transformation implemented by the ideal device from the real one, the discrepancy in the obtained Bayes risk is reflected in the Qloss difference from the magenta experimental points and the continuous curve reported in panel c) of Fig.6.
The control feedbacks chosen by the agent on all the episodes of the training are reported in panel b) of Supplementary Fig.\ref{fig:BayesRisk}.

\begin{figure*}[ht!]
\centering
\includegraphics[width=0.9\textwidth]{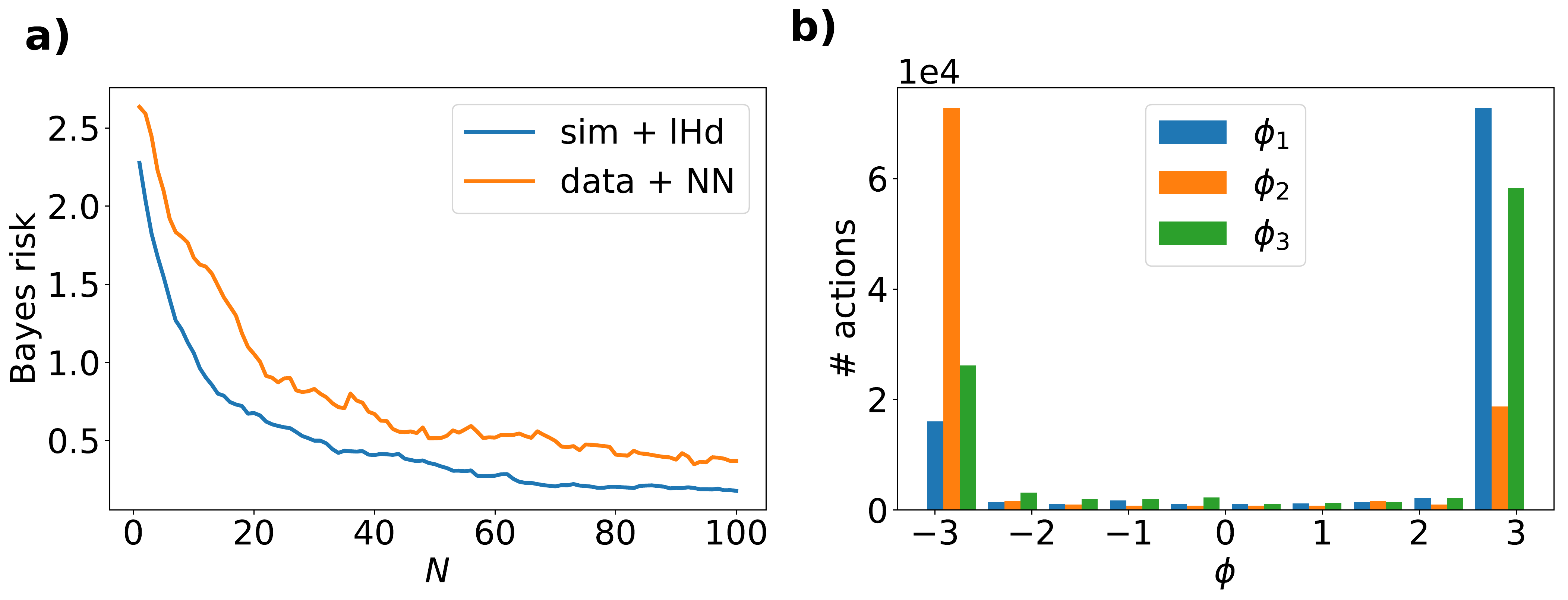}
\caption{\textbf{a)} Bayes risk as a function of the number of sent probes $N$. \textbf{b)} Histogram of different actions imparted by the RL agent for each experiment consisting of $100$ probes. The experiments are performed for $100$ different phases values repated $10$ times each. The three colors refer to feedback imparted on the three different investigated phases.}
\label{fig:BayesRisk}
\end{figure*}
